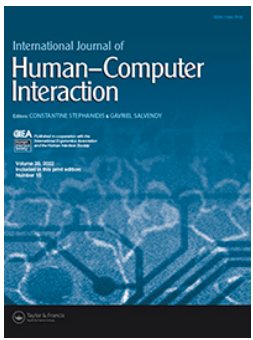

# Wellbeing Supportive Design – Research-Based Guidelines for Supporting Psychological Wellbeing in User Experience

**Dorian Peters**

Published online: 18 Aug 2022.

Submit your article to this journal ⏍

View related articles ⏍

View Crossmark data ⏍

Taylor & Francis
Taylor & Francis Group

# Wellbeing Supportive Design – Research-Based Guidelines for Supporting Psychological Wellbeing in User Experience

Dorian Peters[a,b]

[a]Dyson School of Design Engineering, Imperial College London, London, United Kingdom; [b]Leverhulme Centre for the Future of Intelligence, University of Cambridge, Cambridge, United Kingdom

## ABSTRACT

While human beings have a right to digital experiences that support, rather than diminish, their psychological wellbeing, technology designers lack research-based practices for ensuring psychological needs are met. To help address this gap, we draw on findings from over 30 years of research in psychology (specifically, self-determination theory) that has identified contextual factors shown to support psychological wellbeing. We translate these findings into a list of 15 heuristics and 30 design strategies to provide technology makers with theoretically grounded, research-based, and actionable ways to support wellbeing in user experience.

## 1. Introduction

Growing concerns over the psychological impact of digital technologies are evident in both academic and public discourse, from big tech initiatives on "digital wellbeing" to research in mental health (e.g. Haidt & Allen, 2020; Stiglic & Viner, 2019). While it might be unreasonable to expect all technologies to increase wellbeing, a minimum principle of "do no harm" requires that impact to psychological wellbeing be taken into account during the design process.

Herein we aim to form a bridge between interdisciplinary research and real-world practice in order to help technology designers improve the impacts of their technologies on the psychological wellbeing of users. To do so, we identify, synthesize, and decontextualize knowledge from wellbeing psychology to form a set of evidence-based design guidelines that will help mitigate risk and optimize for positive psychological experience.

Practical rules-of-thumb, otherwise known as "heuristics" or guidelines, have served as research-to-practice tools within technology for decades. The best-known examples are Jakob Nielsen's set of Usability heuristics introduced in the 90s (Nielsen, 1995), Shneiderman's classic "8 Rules of Interface Design" (Shneiderman et al., 2009) and more recently, "18 AI Guidelines for Human-AI interaction" derived by Amershi et al. (2019).

Design Heuristics and guidelines occupy an area between broad *principles* and context-specific design *strategies* or *patterns*. While principles are broad enough to remain always true (e.g. "AI systems should empower human beings") and design patterns are specific to a context (e.g. the pattern for breadcrumbs: "Use linked labels to provide secondary navigation that shows the path from the front to the current site page in the hierarchy"), heuristics and guidelines sit comfortably in between. They are broad enough to be generally true, but specific enough to be actionable and therefore they provide ways of applying broader principles (The Interaction Design Foundation, 2021a). Therefore, a set of heuristics for wellbeing supportive design could help designers systematically account for wellbeing by equipping them with new knowledge for "doing no harm" and providing this knowledge in a familiar form.

While at first blush psychological wellbeing may seem too complex and multifaceted a construct to allow distillation into guidelines. However, wellbeing has been effectively conceptualized, evaluated, and measured for decades by psychologists and economists; wellbeing psychology theories have been applied to technology design (Desmet & Pohlmeyer, 2013; Gaggioli et al., 2017; Peters et al., 2018); and heuristic-style guidelines based on wellbeing psychology already exist within other domains (Martela et al., 2021; Ryan & Deci, 2018).

In this paper, we aim to improve the state of technology design with respect to wellbeing by synthesizing findings from over 30 years of research in psychology (specifically, self-determination theory) to highlight contextual characteristics that support wellbeing. We then transpose these characteristics into a set of 15 heuristics and 30 corresponding strategies (which provide example applications of each heuristic) that can provide designers with theoretically grounded ways to better support wellbeing through design.

CONTACT Dorian Peters  d.peters@imperial.ac.uk  Leverhulme Centre for the Future of Intelligence, University of Cambridge, Cambridge, United Kingdom

## 2. Background and related work

### 2.1. Self-determination theory as a source for wellbeing supportive design guidelines

The human computer interaction (HCI) community draws regularly on research in psychology to improve design practice. Self-determination theory (SDT) – a theory of motivation and wellbeing –has been applied to HCI within various domains (see Peters &Calvo, in press for a review), most extensively within digital gaming (Rigby & Ryan, 2011) but also across domains via the "METUX" model (which stands for "Motivation, Engagement and Thriving in User Experience"). METUX is a holistic practical model for applying SDT across technology design that incorporates evaluation measures and a granular framework for understanding different spheres of technology experience (Peters et al., 2018). While the METUX model applies SDT to technology design, it does not go so far as to identify specific design guidelines or strategies. Therefore, the work described herein can be seen as complimenting or extending work on METUX.

While work on design for wellbeing within HCI also takes other approaches (Desmet & Pohlmeyer, 2013; Hassenzahl et al., 2015; Wiese et al., 2020), SDT is distinctive for its provision of a foundational core – a minimum set of wellbeing requirements that can be applied to *all* technologies, regardless of context or activity. In other words, all designers should, at minimum, ensure that three fundamental psychological needs are met within user experience, because where a design frustrates these needs, there are likely to be negative impacts on wellbeing. SDT defines these three "basic psychological needs" (Ryan & Deci, 2000, 2017) as:

1. Autonomy (a sense of willingness/endorsement, acting in accordance with one's goals and values),
2. Competence (feeling able and effective),
3. Relatedness (feeling connected to and involved with others).

SDT has a mature history of experimental research to support the validity of these needs in a wide range of domains (i.e. health, technology, education, workplace, etc.; see Ryan & Deci, 2017 for a review). Unlike other psychological models of wellbeing, SDT can be applied, not just at the level of life, but also at the finer resolutions relevant to technology design, such as task and interface (Peters et al., 2018) and in ways that can be tested empirically. For example, (Sheldon & Filak, 2008) manipulated all three basic needs via simple variations in wording in a game-learning context and found that supporting all needs improved mood, intrinsic motivation, and game performance.

The advantages of SDT for use in HCI also include its significant evidence base, existing validated measures, and the fact that the three basic psychological needs at the theory's core provide safe targets for design (for more detail, see Peters & Calvo, in press). In addition, research on psychological needs conducted across cultures and age groups provides evidence that these basic needs are essential to healthy functioning universally, even if they are met in different ways and/or valued differently within different contexts (Chen et al., 2015; Church et al., 2013; Yu et al., 2018).

Most pertinent to enabling the work described herein is that SDT research has identified specific (often designable) features of social and technological environments that can satisfy or frustrate psychological needs. It is these features that can inform the development of guidelines.

### 2.2. Existing SDT-derived principles and guidelines

There is precedent for the distillation of SDT-based technology guidelines within specific application domains. For example, Van Roy derived SDT-based heuristics for gamification in digital education (van Roy & Zaman, 2017), while Yang and Aurisicchio (2021) derived SDT guidelines for conversational agents. Szalma (2014) presented SDT concepts as broad principles for supporting autonomous motivation in human factors, which the author hoped would help provide a basis for the development of more practical design recommendations. Our objective herein is to take a domain-agnostic approach and derive actionable heuristics for technology design that are as broadly applicable as usability guidelines. This process we used is detailed below.

### 2.3. Method: Moving from characteristics to heuristics

The impact that design decisions can have on psychological needs is evident in findings across domains that identify which characteristics of environments support needs. For example, within the SDT literature on psychotherapy, clinicians are advised to "practice empathic listening" and to "provide a meaningful rationale for therapeutic strategies" as these have consistently proven effective at fulfilling the three psychological needs. Likewise, educators are advised to support autonomy by "providing meaningful choices" (Lonsdale et al., 2009) while sport coaches can support autonomy by fostering intrinsic rather than extrinsic goals (Ingledew & Markland, 2008).

The research in SDT is spread widely across disciplinary literatures ranging from education, marketing and sport to long-haul spaceflight and hip-hop dancing (e.g. Carpentier & Mageau, 2013; Goemaere et al., 2016; Kim & Drumwright, 2016; Niemiec & Ryan, 2009; Quested & Duda, 2009). This makes any attempt at a comprehensive review particularly challenging. However, the production of handbooks that summarize work in SDT provide consolidated sources for need-supportive characteristics. As such, the seminal and comprehensive text, *Self-Determination Theory: Basic Psychological Needs in Motivation, Development, and Wellness* (Ryan & Deci, 2017) was selected as the primary referential source for the work synthesized herein. This was supplemented by a review of SDT work in technology (Peters & Calvo, in press), and three existing consolidated lists of SDT-based guidelines for specific contexts: coaching (Ryan & Deci, 2018), behaviour change (Teixeira et al., 2020) and crisis communication (Martela et al., 2021).

Among the characteristics that support need satisfaction, we have only included those that can be applied to technology design with minimal adaptation. For example, behaviour change recommendations include "show unconditional regard" and "Encourage asking of questions" (Teixeira et al., 2020) which do not seamlessly translate to technology design and so were not included herein. By contrast, the recommendations: "assist in setting optimal challenge", "offer constructive, clear, and relevant feedback" (Teixeira et al., 2020) and "use non-controlling, informational language" (Martela et al., 2021) do readily apply to technology contexts and were therefore included (see heuristics 7, 10, and 12 below). We have formulated those included into a format and structure that facilitates their application to technology design (Table 1). We also provide technology-specific examples for each heuristic as further evidence of applicability.

**Table 1.** Heuristics for wellbeing supportive design which identify research-based techniques for satisfying basic psychological needs.

| 15 Heuristics for wellbeing supportive design |
| --- |
| **Principle 1: support autonomy** |
| 1 Apply best practices for accessibility |
|   Ensure the technology can be accessed by all end users regardless of ability. |
| 2 Empathize with your user's frame of reference |
|   Seek to listen to, and understand, your users' feelings, perceptions, values, and goals. |
| 3 Provide meaningful rationale |
|   Provide meaningful rationale for requests, suggestions and system actions, in order to support endorsement. |
| 4 Provide meaningful choices |
|   Provide the user with choices that are of value to them. |
| 5 Support informed choices |
|   Provide the user with information necessary for making choices. |
| 6 Communicate in autonomy-supportive, rather than controlling ways |
|   Use language and communication styles that support volitional action rather than coercion or control. |
| 7 Ensure rewards are autonomy-supportive |
|   Provide informational rewards that function, not as ends in themselves, but as feedback that is linked to valued behaviours, rather than as contingencies that apply pressure. |
| 8 Support mindful attention |
|   Design to support volitional experiences of focus, mindful awareness and attention. |
| **Principle 2: support competence** |
| 9 Apply best practices for usability |
|   Follow established usability guidelines to ensure functionality and ease-of-use. |
| 10 Support optimal challenge |
|   Ensure the level of complexity or challenge is appropriate for the user and context. |
| 11 Provide non-evaluative feedback |
|   Provide feedback on progress, effort, and specific behaviours rather than assessing the person. |
| 12 Provide effectance-relevant feedback |
|   Provide feedback on controllable features and informs a pathway to improvement. |
| **Principle 3: support relatedness** |
| 13 Support a sense of meaningful connection to others |
|   Support experiences of warm connection and closeness to others. |
| 14 Support a sense of belonging |
|   Support experiences of significance and belonging to larger social contexts. |
| 15 Support caring for others |
|   Facilitate kindness and opportunities to care for others |

The heuristics presented are formulated according to the following criteria based on Amershi et al. (2019) which dictate that they:

- should be written as a rule of action, containing about 3–10 words and starting with a verb,
- should be accompanied by a one-sentence description that qualifies or clarifies any potential ambiguities,
- should not contain conjunctions so that designers can clearly validate whether it is applied or violated in an interface.

We add that each is presented alongside specific references to research evidence for validity. We have also opted to present each heuristic together with specific corresponding design strategies (30 in total) which serve to provide examples of how these heuristics can be applied. In addition, two external researchers (one in psychology and one in HCI), both with extensive experience applying self-determination theory, provided feedback and agreement on the final guidelines providing additional confidence with respect to theoretical and practical integrity.

## 3. Principles, heuristics and strategies

### 3.1. Introduction

Below are 15 heuristics for supporting the three basic psychological needs of autonomy, competence, and relatedness in technology design. The heuristics are summarized in Table 1 and described in detail (with research evidence and examples) within this section. Note that a strict non-maleficence ("Do no harm") approach would only seek to ensure none of these heuristics were violated. A beneficence approach would seek opportunities to apply them more proactively to design. Some will apply more readily than others for a given technology or project. The heuristics and strategies have been categorized according to the basic psychological need they address as expressed by three basic *wellbeing supportive design principles*:

1. Support autonomy
2. Support competence
3. Support relatedness

It should be noted that psychological needs are inter-dependent and interactional. In general, existing evidence shows they facilitate each other, so that providing support for one can increase others (Ryan & Deci, 2017, pp. 167, 311). As such, some heuristics and strategies classified under one psychological need may also serve others.

Finally, the heuristics and strategies presented below have also been provided as a card deck, and part of a wellbeing supportive design toolkit, which is freely available at http://www.positivecomputing.org.

## 3.2. Support autonomy

To support autonomy entails supporting people to act *willingly*, in ways they *endorse*, and in accordance with their goals and values (Ryan & Deci, 2017). When users can't influence an interface in alignment with their goals, they get frustrated and their sense of autonomy is undermined. It is therefore autonomy that lies at the heart of many usability guidelines, e.g. "User control and freedom" (Nielsen, 1995). It's worth noting that autonomy is often erroneously confused with independence *per se* (i.e. being self-sufficient), or with "being in control". However, one can willingly embrace *inter*dependence (e.g. as part of a family or community) or happily relinquish control (e.g. to a trusted authority) without diminishing one's sense of autonomy. Therefore, when people act autonomously, they are, above all, self-endorsing this action. It is autonomy's link to goals and values that connects it to a sense of meaning & purpose in our lives.

### 3.2.1. Heuristic 1: Apply best practices for accessibility
#### 3.2.1.1. Definition. Ensure the technology can be accessed by all intended users.

#### 3.2.1.2. Description. A technology that cannot be accessed cannot support autonomy. Universal design refers to "the design and composition of an environment so that it can be accessed, understood and used to the greatest extent possible by all people regardless of their age, size, ability or disability" (National Disability Authority, 2020) and therefore constitutes another foundation for need support.

#### 3.2.1.3. Strategies.
- S1. *Comply with web accessibility standards* – The Web Content Accessibility Guidelines (WCAG) are a stable technical standard (ISO/IEC 40500) Composed of 14 guidelines or "success criteria" for web and mobile content.
- S2. *Test with a variety of platforms and with a range of users* - Thorough testing, and anticipation of edge cases, can help pre-empt autonomy frustrations arising from the wide diversity of real-life constraints within which users access technologies.

### 3.2.2. Heuristic 2: Empathize with your user's frame of reference
#### 3.2.2.1. Definition. Seek to listen to, understand, and design for your users' feelings, perceptions, values, and goals.

#### 3.2.2.2. Rationale. You can only support user goals and values if you know what they are. The notion of a "frame of reference" refers to a person's feelings, perceptions, values, and goals, as well as potential conflicts and barriers (Ryan & Deci, 2017, p. 443). The notion of user empathy is well-established within user experience practice which provides research methods for listening to users in order to better understand their needs and perspectives. Moreover, empathic listening can continue as part of the product or service once it is deployed. For example, a technology might check in on changing preferences, learn from user data (ethically and with consent), or provide opportunities for feedback and customization. This heuristic relies on a mindset: genuine interest in supporting the user to function autonomously. Such a mindset may not always align with all business models but does align with positive psychological outcomes.

#### 3.2.2.3. Strategies.
- S3. *Use qualitative research and human-centered methods* - Participatory design, design thinking, and human-centred design (IDEO.org, 2015; Sanders & Stappers, 2014; The Interaction Design Foundation, 2021b) all provide methods for better understanding the user's frame of reference (co-design workshops, interviews, ethnography, etc.), and for translating this understanding into artefacts that can help the larger team empathize during development (e.g. personas, user stories, etc.) Just as HCD methods can help designers identify ways to fulfil psychological needs, psychological needs can help add evidence-based specificity and focus to HCD efforts and provide a way to attend to wellbeing explicitly and in a way that can be measured.
- S4. *Provide ongoing opportunities for user feedback and input* - People's perspectives, orientations, and needs are in constant flux. Therefore, design can benefit from information on changing needs over time (especially when a technology is designed to facilitate such change, as in learning or health).

### 3.2.3. Heuristic 3: Provide meaningful rationale
#### 3.2.3.1. Definition. Provide meaningful rationale for requests, suggestions, and system actions in order to support endorsement.

#### 3.2.3.2. Rationale. A person doesn't always need to be in control to feel autonomous, but they do need to feel they endorse their actions. This relies on understanding the rationale for actions they might take (Ryan & Deci, 2018). In the context of therapy, this has been described as "a rationale that is tailored or personally meaningful for any activities they are to engage in." (Ryan & Deci, 2017, p. 444). In a technology setting, this might involve rationale for any requests or suggested options – for example, an

explanation for why the system is requesting personal data; why it's worth engaging in a particular task; or for why the system has to load, update, or otherwise interrupt the task flow. "Meaningful" is an important qualifier. Users don't generally want detailed technical explanations of every system event. People don't tend to "endorse" extraneous information overload. Therefore, it's important to ensure the rationale provides information that is relevant to the user and at points in which a situation would otherwise cause need frustration.

### 3.2.3.3. Strategies.
- *S5. Provide optional levels of explanatory detail* - Different people will want varying degrees of explanation with respect to rationale. Some will have less trust in a system, in the organization behind it, or in technology generally, and will therefore require more rationale in order to endorse something. Therefore, allowing users to drill down into more detail as desired ensures sufficient rationale is available for those who want it without overwhelming those who don't. The commonly used "Why are we asking for this?" link that is provided beside form fields is one example of this approach.

### 3.2.4. Heuristic 4: Provide meaningful choices
#### 3.2.4.1. Definition. Provide the user with choices that are of value to them.

#### 3.2.4.2. Rationale. Critical to autonomy is the feeling that one has influence over one's actions and environment. In the digital environment, this will often come in the form of options for changing that environment to better suit one's goals and values. Even within inevitable constraints, it's important to indicate which choices people still have (Martela et al., 2021). Again, the term "meaningful" is crucial. Providing more choices *per se*, "when none of the options has real value to the person or when there are so many options to choose from that the process becomes burdensome (e.g. Iyengar & Lepper, 2000)" does not increase autonomy and risks infringing on competence. User research should help direct design toward choices that are valued.

#### 3.2.4.3. Strategies.
- *S6. Allow users choice over goals and strategies* - Referred to as Opportunities for Action (OFAs) in game environments, choice over goals and strategies are recognized as critical to enjoyable play (Ryan & Deci, 2017, p. 516). In behaviour change, opportunities to set one's own goal (e.g. a health goal) as well as to select one's own parameters, measures of success, and approaches to getting there (collectively known as implementation planning) have been shown to increase the chances of reaching goals (Ryan & Deci, 2017, p. 449).
- *S7. Offer customizable profiles* - Customization of profiles or avatars can increase experiences of autonomy by supporting a sense of personal relevance and agency within a virtual environment (Ryan & Deci, 2017, p. 515). To illustrate Peng et al. (2012) found that the presence of avatar personalisation in an exercise game significantly improved ratings of autonomy, which in turn predicted higher enjoyment, motivation for future play, and overall ratings.

### 3.2.5. Heuristic 5: Support informed choices
#### 3.2.5.1. Definition. Provide the user with information necessary for making choices.

#### 3.2.5.2. Rationale. Critical to "true choice" according to SDT is that choice is *informed*. If a design offers choices but withholds critical information required to select from those choices, then it does not support autonomy. In the context of coaching, Ryan and Deci (2018) explain that "awareness is what allows people to make a decision that is a true choice … facilitating choice means providing the support that allows clients to home in on what they value, and choose actions congruent with that awareness". A salient example of the violation of this heuristic is the lack of informed consent that notoriously occurs with many end-user license agreements. In theory, we make a choice to consent to questionable uses of our data, when in reality, the agreements are so lengthy and difficult to understand that users seldom make that decision autonomously. Recent improvements in making these agreements more clear and understandable represent advancement toward better support for autonomy.

#### 3.2.5.3. Strategies.
- *S8. Facilitate safe exploration and experimentation* - One way to inform choice, is to allow users to safely try out options beforehand. Open world games present the epitome of safe exploration as players do what they please with agency in a low-consequence environment. However, the opportunity to try things out can also be found in learning (as with learning through models and simulations) and as an approach to AI transparency (Shneiderman et al., 2009). At the most basic level, allowing users to preview what would happen if they made a particular choice – confident that they can undo or change their mind without cost – has also proven effective for product onboarding as the "free trial". This can help support autonomous decision-making so long as it doesn't involve traps like the "roach motel" (a forced subscription that is easy to get into but hard to get out of) (Brignall, 2021).
- *S9. Allow users to agree selectively* - While systems in the past often forced users to agree to an opaque all-or-nothing deal in exchange for a service, many software systems are now giving users finer grained control. For example, Apple's iOS allows users to selectively agree to give an app access to just "some photos" rather than the whole library. Similarly, video conferencing allows the sharing of just audio or just video; and many organizations allow users to select which kinds of subscription emails they are happy to receive.

### 3.2.6. Heuristic 6: Communicate in autonomy-supportive rather than controlling ways

#### 3.2.6.1. Definition. Use language and communication styles that support volitional action rather than coercion or control.

#### 3.2.6.2. Rationale. In general, people perceive controlling language as autonomy frustrating which proves detrimental to both engagement and wellbeing (Teixeira et al., 2020). As Ryan and Deci (2017, p. 448) put it in the context of behaviour change: "Having feedback and guidance, rather than control or directives, can help a person feel more purposive and confident in engaging in potential change." While there are a few situations in which people may endorse commanding language (e.g. when preventing immediate risk) under most circumstances, a non-judgmental communication style that supports user decision-making will best support autonomy. For more detail see the set of principles for autonomy-supportive language in crisis communication (Martela et al., 2021).

#### 3.2.6.3. Strategies.

- **S10. Provide feedback and guidance rather than directives** - Research shows that in supporting progress towards a goal (such as with learning or behaviour change technologies) language that guides rather than directs will improve outcomes. For example, in the context of coaching, Johnson et al. (2022) explain that autonomy-supportive communication can "be inviting and informational ... (e.g. "I propose," "you could"), attempt to reduce pressure (e.g. through avoiding coercion, emphasis of reward, obligation, or guilt), and avoiding controlling language (e.g. "you must" "do not")."

- **S11. Support reflection** - Designing technology to support reflection (rather than merely provide an evaluation) has at least two benefits: 1. it supports autonomy and competence by avoiding directives, and 2. it avoids need-frustrating scenarios in which the technology makes incorrect assumptions. For example, in the development of *EQ Clinic* a communication training tool for medical students, automated data on various features of remote interviews with patients were provided as feedback to medical students for reflection (e.g. data on turn-taking: "you were talking about 70% of the time"). This allowed students to reflect on their communication style, for example, by asking, "Am I dominating the conversation?". The system designers deliberately opted *not* to provide evaluative feedback (e.g. "You're talking too much") as the student and teacher are in better positions than the system to interpret very contextually-dependent data, and the act of reflecting and interpreting is itself useful to learning (Kori et al., 2014).

- **S12. Provide controls over notifications and communications** - Even non-controlling feedback can frustrate autonomy and feel coercive if it's too frequent or comes at the wrong time. Making frequency and type of feedback customizable can help prevent this. The notification controls provided by Apple's iOS provides a good example of highly customizable notification preferences.

### 3.2.7. Heuristic 7: Ensure rewards are autonomy-supportive

#### 3.2.7.1. Definition. Provide informational rewards that function, not as ends in themselves, but as feedback that is linked to valued behaviours, rather than as contingencies that apply pressure.

#### 3.2.7.2. Rationale. The use of punishments, or even positive reinforcements as incentives, can apply undue pressure thus frustrating autonomy. Furthermore, rewards (e.g. prizes and badges) when too strictly contingent on performance have proven demotivating (Ryan & Deci, 2017, pp. 123–157). This doesn't mean there is no place for rewards and incentives, but that they need to be designed with care. Specifically, it's important to ensure the reward is attached to the right behaviour rather than to a proxy or outcome (which could encourage users to game the system by finding shorter or unethical paths to the outcome)(Ryan & Deci, 2017, pp. 142–143). It can be helpful to think of rewards, either as feedback on progress (e.g. a badge for reaching 100k) or as surprise delighters (reinforcing an achievement), rather than as elements for pressuring the user to reach a goal. The key is that users are not taking action primarily to get the reward.

#### 3.2.7.3. Strategies.

- **S13. Focus on process rather than outcomes.** - SDT suggests "a focus on process rather than outcomes" (Ryan & Deci, 2019) including for rewards. This is because informational rewards, for example, those that reflect effort and progress are more likely to be experienced as supportive of autonomy and competence. In contrast, rewards contingent on engagement or performance are more likely to be perceived as controlling, i.e. as pressures to think, feel or behave in particular ways (Ryan & Deci, 2017, pp. 130, 546).

### 3.2.8. Heuristic 8: Support mindful attention

#### 3.2.8.1. Definition. Design to support volitional experiences of focus, mindful awareness and attention.

#### 3.2.8.2. Rationale. One of the most widely-voiced concerns about digital experience is distraction or "attention hijacking" – the interruption of volitional attention made by technologies, usually for commercially-driven purposes such as "driving engagement" elsewhere. The efficacy and frequency of this hijacking has been amplified by advancements in AI, such as microtargeting and recommender systems (Calvo et al., 2020; Wu, 2017; Zuboff, 2015). When human attention is for sale, and technologies can "push" advertising, alerts, and clickbait into our consciousness unbidden through exploitation of primal attention grabbers, then our autonomy can be compromised by an inability to

maintain self-direction. Even, simpler aspects of design, such as a poorly crafted interface, can hinder our ability to stay on task or maintain focus. The removal of distractions, extraneous cognitive load, and interruptions will help prevent this autonomy frustration. In addition, SDT has long argued for the strong role of awareness in autonomy, and for the role of mindfulness practice in cultivating such awareness. "Mindfulness is defined as the open, receptive awareness of what is occurring (Brown & Ryan, 2003). It allows people to contact information from both internal sources (perceptions, feelings, and values) and external events, and to use this information to come to a clear focus and gain a sense of what, all things considered, one would most value doing." The power of mindfulness to support both autonomy and wellbeing has been demonstrated by numerous studies, including within SDT research (Brown et al., 2015; Brown & Ryan, 2003; Ryan & Deci, 2018). This heuristic may not always align with all business models, but it does align with positive psychological outcomes.

### 3.2.8.3. Strategies.

- **S14. Simplify the interface to support focus** - Extraneous design features (e.g. too many colours, decorative graphics or movement), even when well-intended, can disrupt attention and hinder focus. Therefore, following guidelines for the reduction of extraneous cognitive load can support attention and help users maintain it. For detailed strategies, refer to related work on design for learning (Peters, 2014).

- **S15. Minimize interruptions** - Different technologies have taken various approaches to help people reclaim and retain autonomy over their attention. For example, Apple's iOS controls allow for the silencing of notifications, while Microsoft Word's "focus view" transforms the interface into a simple view of the bare essentials necessary for writing. Indeed, a whole genre of software has emerged to meet the user demand for ways to apply constraints on their own access to services such as social media or the internet, in order to reclaim their digital environments for undivided attention. Allowing users nuanced control (where, when, how and how often) over those aspects of digital experience that might interrupt or distract them can help balance autonomy with a need to be kept informed.

### 3.2.9. Diagnosing autonomy frustration

On behalf of facilitating a "do no harm" baseline in which designers, at minimum, avoid *frustration* of basic psychological needs, we can extrapolate from the recommendations above to identify a number of signs and symptoms of autonomy frustration within digital experience. Namely, experiences of autonomy frustration may feel:

- Controlling
- Manipulative
- Intrusive
- Nagging
- Inflexible

- Evaluative/judgmental
- Non-consensual

These signs can be attended to for diagnostic purposes. When users describe technology use in ways that reference elements such as these, designers can understand this experience in terms of autonomy frustration and look for ways to improve autonomy

### 3.3. Support competence

Competence is defined as feeling capable and effective and involves the intrinsic drive for self-efficacy, growth, learning and mastery. Therefore, **"**feelings of competence come about when people have opportunities to apply skills and effort to tasks that are moderately difficult, allowing them to experience efficacy and success and thus to derive feelings of mastery and competence." (Ryan & Deci, 2017, p. 513). Thus, factors that have been shown to enhance a person's sense of competence include optimal challenge, positive feedback, and opportunities for learning – elements familiar to both game designers and educators. Even when someone is not engaged in learning, competence frustrations emerge when they feel incapable or ineffective. When a user stares bewildered at an overly complex screen, their competence is taking a hit. Usability heuristics such as "recognition rather than recall" and "help and documentation" are important for competence support (Nielsen, 1995).

### 3.3.1. Heuristic 9: Apply best practices for usability

#### 3.3.1.1. Definition. Follow established usability guidelines to ensure functionality and ease-of-use.

#### 3.3.1.2. Rationale. While it may sound overly obvious to seasoned designers, it is important to note that all usability guidelines are protections of autonomy and competence needs. Therefore, ensuring usability is foundational to wellbeing support, because if a technology is not usable, it will frustrate psychological needs.

#### 3.3.1.3. Strategies.

- **S16. Conduct usability testing** - It probably shouldn't come as a surprise that what is good for usability is good for psychological experience. This makes usability testing important to a "do no harm" principle. Even a technology designed to provide a psychological treatment could undermine its goals by neglecting usability.

### 3.3.2. Heuristic 10: Support optimal challenge

#### 3.3.2.1. Definition. Ensure the level of complexity or challenge is appropriate for the user and context.

#### 3.3.2.2. Rationale. There are many important contexts in which challenge is intrinsic to a digital experience in productive ways (e.g. behaviour change, education, gaming). In these cases, intrinsic challenge must neither be too hard nor too easy in order to support competence. In education, challenge

is inherent to the learning process, but is carefully adjusted for the learner's level through techniques like scaffolding and chunking (Niemiec & Ryan, 2009). Similar approaches are taken in gaming to optimize enjoyment, and high competence frustration in games has been linked to short-term post-game aggression. In behaviour change, "Optimal in SDT means challenges that are readily but not easily mastered and that are not overly stressful or demanding." Of course, there are many circumstances in which challenge is *not* an intrinsic or productive element of activity (e.g. online shopping). In these cases, simply preventing obstacles and minimizing difficulty becomes the focus. Anything instrumental, such as a user interface, should not pose challenges itself.

### 3.3.2.3. Strategies.

- *S17. Break down big tasks into manageable parts* - In education, this is called "chunking" or "segmenting", and in behaviour change, it has sometimes been framed as breaking behaviours down into "tiny" habits (Fogg, 2019). Breaking things down has two benefits: creating more proximal goals that are more immediately achievable and providing a string of ongoing competence satisfaction experiences. For example, tax software relieves a major competence frustration by walking the user through an easier version of what is normally an overwhelming (and taxing) process. The software makes it more manageable through methods like chunking, sequencing, plain language and scaffolding. Complexity can also be managed with techniques like progressive disclosure and layering (Peters, 2014) as well as through the provision of structure defined in SDT as "clarity of goals and scaffolding of tasks such that success is an ongoing proximal outcome, and these proximal successes build toward distal goals and accomplishments" (Ryan & Deci, 2018).

- *S18. Provide adaptation and levelling* - Users change as they use technologies, and not merely within learning and behaviour change contexts. Therefore, it makes sense that technologies should be able to change with them. Strategies for adapting to changing needs include progressive disclosure (Babich, 2020) and levelling in gaming (Ryan & Deci, 2017, p. 514). For example, in an exergame experiment (Peng et al., 2012) tested a version of the game with dynamic difficulty and showed that this adaptation not only resulted in better game enjoyment, it did so via greater competence satisfaction.

- *S19. Make challenge customizable* - While games can readily adapt to users' changing expertise, technologies don't always have the advantage of continual performance feedback. Therefore, allowing users themselves to optimize difficulty can be helpful. For example, in the behaviour change arena, some diet apps allow users to choose their goals, strategies and difficulty level (i.e. a custom timeframe for achieving a goal or "easy", "medium" and "hard" versions of diets).

### 3.3.3. Heuristic 11: Provide non-evaluative feedback
#### 3.3.3.1. Definition. Provide informational feedback on progress, effort, and specific behaviours, not on the person.

#### 3.3.3.2. Rationale. Feedback is most effective when it provides encouraging information on progress or effort with respect to a specific goal or behaviour, as opposed to feedback that marks something as "good" or "bad", or that implies a judgement of fixed traits or self-worth (Teixeira et al., 2020). In other words, feedback should be about behaviours, rather than about the person themselves. Furthermore, lack of progress should be treated empathically and informationally rather than evaluatively, i.e. as an opportunity to learn about obstacles arising and to consider strategies in response. This approach applies equally to support for self-monitoring: "Done in a way that is autonomy-supportive toward oneself, self-monitoring is not self-evaluative – it is instead self-informative."(Ryan & Deci, 2018)

#### 3.3.3.3. Strategies.

- *S20. Provide informational feedback* – Informational feedback provides insights on behaviour without pressure for a particular outcome and without adding to a feeling of being evaluated, judged or surveilled (Ryan & Deci, 2017, 2018, p. 148). Moreover, positive feedback with evidence of progress (without pressure), is more likely to be perceived as autonomy-supportive (Hagger et al., 2015). Some activities naturally and intrinsically come with informational progress feedback (e.g. as you climb a mountain, there's visual feedback on how far you've gone) but technologies can especially help in contexts where this task-inherent feedback isn't available by providing other ways for people to gauge their progress.

### 3.3.4. Heuristic 12: Offer effectance-relevant feedback
#### 3.3.4.1. Definition. Provide feedback on controllable features and that informs a pathway to improvement.

#### 3.3.4.2. Rationale. Feedback is most effective when it is clear what it relates to and carries information on strategies for improvement. This "Effectance-relevant" feedback, is also called "constructive" or "instructional" feedback (Fong et al., 2019) and targets controllable features of performance (Carpentier & Mageau, 2013). As elaborated within the coaching context: "Competence supports include communication of structure, strategy options, feedback, and clarity of limits" (Ryan & Deci, 2018). Constructive feedback is also most effective when given immediately, or as close in proximity as possible to the behaviour to which it refers.

#### 3.3.4.3. Strategies.

- *S21. Provide proximal and multi-level feedback* - In real life, the pathway to success can be ambiguous, feel distant, and have few rewards along the way. Technologies can help by supporting users in clarifying goals and providing feedback more frequently along the path and in ways that help them recover from setbacks and adjust to changing needs. Rich feedback mechanisms are nowhere more salient than in digital games. Rigby and Ryan (Rigby & Ryan, 2011) highlight three types: 1. tightly-coupled and immediate *granular feedback* on effectance (as in sound explosions when you hit the right button); *sustained feedback* (as in the increasingly roaring crowd over time in dance games, or "streaks" in some fitness apps); and *cumulative feedback* which charts progress over time and might lead to unlocking new levels or other evidence of moving towards a goal.

### 3.3.5. Diagnosing competence frustration

We can extrapolate from the recommendations above to identify a number of signs and symptoms of competence frustration within digital experience. Namely, experiences of competence frustration may feel:

- Too difficult
- Clumsy
- Overly complex
- Overwhelming
- Hard to follow
- Demoralizing
- Defeating
- Stagnant or boring
- Unmanageable in available time or with available resources/skills.

When users describe technology use in ways that reference elements such as these, designers can understand this experience in terms of competence frustration and look for ways to improve competence support.

### 3.4. Support relatedness

Relatedness is described as a sense of belonging and connectedness to others and is central across wellbeing theories. "Relatedness needs are satisfied when others recognise and support one's self and when the person feels able to connect with, feel significant with, and be helpful to others." Therefore, relatedness can be fulfilled through prosociality, such as acts of giving, as well as through opportunities to feel cared for (Ryan & Deci, 2017, p. 11). Technologies increasingly support social connection, however, not all social interaction (technologically-mediated or otherwise) helps us feel a greater sense of connectedness. From collecting "friends" and performing on social media ("me-forming"), to comparing oneself to idealized others, and exposure to divisive dis-information and the social echo-chambers that result, there are many ways in which digital social connection does not fulfil the need for relatedness and may even hinder it.

Relatedness is different to autonomy and competence needs within the technology context in that it is not *always* relevant to design. This is because, while relatedness is essential to life and important to many technologies, not *all* technologies will have impact on relatedness experience. For this reason, this section has more strategies than heuristics, as context-specificity is greater. Nevertheless, for a "do no harm" principle, it remains critical to understand relatedness needs in order to avoid diminishing them. To this end, it is always important to consider how a user will engage with a technology in the context of their lives which includes their relationships with community and others. This is even true for technologies without any explicit social features.

### 3.4.1. Heuristic 13 – support a sense of meaningful connection to others

#### 3.4.1.1. Definition. Support experiences of warm connection and closeness to others.

#### 3.4.1.2. Rationale. Relatedness involves feeling meaningfully connected to others. Social features can provide support for this connection but have ramifications that need to be thought through, tested, and made customisable in order to ensure support for genuine relatedness and other psychological needs. For example, studies on Facebook show that directed communication has more positive psychological impact while mere browsing can be negative (Burke et al., 2010). Insights like these, which shed more nuanced light on digital social interaction, can inform design – in this case by suggesting "potential design enhancements for fostering communication over passive engagement." (ibid). Similarly, competition can be either motivating or disconnecting (at worst, it can even feed aggression(Adachi & Willoughby, 2011). Therefore, the addition of social features must always be carefully assessed. On the flip side, technologies can support deeper social connection without social features at all. An effective gratitude journal or mindfulness app, for instance, can foster reflection or mindful attention that enhances a person's sense of connection with others without requiring technology-mediated contact. Finally, meaningful connection need not be limited to other humans. SDT research has demonstrated that contact with nature is both conducive to, and directly fulfilling of, our need for relatedness (Ryan & Deci, 2017, p. 265). Exposure to nature has a positive effect on autonomy as well and promotes prosocial behaviour and "a focus on intrinsic values for social relationships and community rather than on personal gain" (even just natural imagery or potted plants have produced results, although more immersion is better).

#### 3.4.1.3. Strategies.

- *S22. Ensure technology-mediated interactions with others are seamless* – Remarkably, the mere presence of a smartphone has been shown to decrease connection quality in face-to-face conversations (Dwyer et al., 2018). Even a small lag on a video call can diminish one's sense of connection to the other person on the call. Technologies can interrupt and diminish close connection through distraction or poor performance. Therefore, designers should seek to disrupt social engagement as minimally as possible and to facilitate smooth, intuitive and responsive social experiences. Technology social features should also be customizable to allow participants to tailor disclosure and disruptions to the context.
- *S23. Provide opportunities to connect with close others both offline and online* – Adults report that a major reason for playing games is to connect with other people, that relatedness is associated with a sense of meaning in games, and that multiplayer options can increase relatedness satisfactions (Ryan & Deci, 2017, p. 518). In response to the COVID-19 lockdown, people began gathering virtually in new ways, for example, as avatars in 3D space, at virtual house parties, and with myriad whiteboards, breakout rooms, and video meetings across the globe. Technologies can also facilitate the organization of in-person (non-virtual) meet-ups and activities.

- *S24. Focus feedback on intrinsic vs extrinsic relatedness goals* – SDT research has shown that not all goals are equal. Some goals are likely to fulfil basic psychological needs and promote wellbeing, whereas others, even when achieved, are likely to frustrate basic needs, and yield ill-being (Sheldon et al., 2004). Specifically, researchers have found that extrinsic (instrumental) life goals (such as when people make it highly important to accrue wealth, present an attractive image, or become popular) do not satisfy psychological needs whereas intrinsic goals (valuable for their own sake) such as growth, intimacy, and service to others (Sheldon et al., 2004; Sheldon & Krieger, 2014) do. Thus, designers can attend to the goals encouraged by technologies and to the motivations behind those goals. For example, pushing users to increase followers, likes, or other status symbols is unlikely to support relatedness as effectively as supporting goals to, for example, help others or connect more deeply.

- *S25. Support contact with nature and non-human life* – Although technologies are often associated with disconnection to nature, finding creative ways to increase this connection can support relatedness (and autonomy too). Some technologies support outdoor interaction, for example, by incorporating outdoor activity in gameplay or by providing access to information that is physically situated (e.g. apps that recognize plants, provide informative tours, or support outdoor recreation.) Even virtual experiences of nature have shown benefits to wellbeing (Grinde & Patil, 2009) and the popularity of relaxation apps that provide nature sounds and visuals are likely a testament to this. Moreover, although, SDT research has yet been conducted on the relatedness fulfilment of human-animal interaction, such a hypothesis is certainly aligned with research in other fields (McConnell et al., 2019; Wells, 2009).

### 3.4.2. Heuristic 14: Support a sense of belonging
#### 3.4.2.1. Definition. Support experiences of significance and belonging to larger social contexts.

#### 3.4.2.2. Rationale. Feeling a significant part of something larger than oneself can involve feeling part of a family, community, society, social endeavour, planet, etc. Technology has provided many new opportunities for humans to connect with larger groups and communities from the local to global in scale. Feeling belongingness involves feeling significant to those you care for which can be nourished by signs of appreciation, praise and gratitude. Moreover, technologies can help make the presence of others salient virtually even when individual users are physically isolated and might otherwise feel alienated or excluded.

#### 3.4.2.3. Strategies.
- *S26. Make community salient* - A sense of belonging and human connection can often be facilitated in a virtual environment by making social presence more salient. This can be done in the abstract (e.g. "You are part of a community of 10,000 others nationally who also love to knit") or literally (e.g. "284 other knitters are currently online"). However, be aware of saliency that feels like manipulation (e.g. "23 other customers are viewing this"), pressure (e.g. read receipts) or surveillance, which can be frustrating to both relatedness and autonomy. This kind of salience is not providing a sense of meaningful belonging. Providing choice over individual salience (e.g. "Show as offline", "Don't show when a message has been read") can provide important safeguards.

- *S27. Provide opportunities to show appreciation and gratitude* - Feeling significant to those you care about is part of feeling you belong. Technologies can help make this significance more salient by providing simple ways for people to express appreciation and gratitude.

### 3.4.3. Heuristic 15: Support caring for others
#### 3.4.3.1. Definition. Facilitate kindness and opportunities to care for others.

#### 3.4.3.2. Rationale. Relatedness involves feeling cared for, but also caring for others. This includes having opportunities for benevolence and beneficence (i.e. having a positive impact on others)(Martela & Ryan, 2020). Indeed, technology has provided many new opportunities for humans to share, help, and otherwise contribute prosocially to community endeavours on both intimate and global scales, from the incidental (e.g. sharing inspiring videos or in-game rewards), to the global (e.g. micro-lending and crowdsourced sharing during natural disasters).

- *S28. Provide opportunities to contribute or share with others* - Many technology-mediated contexts provide support for members of communities to help each other, to cooperate towards a shared goal or contribute to a social good. Technologies have made it easier to recycle, share food, crowdfund, share knowledge, and provide moral support allowing experiences of giving and caring for others.

- *S29. Make positive impact salient* - Receiving evidence that we have positively impacted others can increase both relatedness and competence. For example, many digital services that rely on volunteer contributions provide evidence of positive impact (e.g. "33 people were helped by your review") while charities increasingly provide specific feedback on outcomes (e.g. "Your donation provided 50 meals to children in need".) Simply allowing users to easily express gratitude is another approach. For example, Naqshbandi et al. (2020) showed that the expression of gratitude by the beneficiary of volunteering affected the volunteer's experience of relatedness which then correlated with increased volunteering.

- *S30. Provide opportunities to offer kindness and praise* – Sometimes caring is experienced and expressed through

the sharing of kind words. Technologies can facilitate the sharing of kind words such as support, congratulations and praise. For example, some workplace social networks scaffold the provision of praise.

### 3.4.4. Diagnosing relatedness frustration

We can extrapolate from the recommendations above to identify a number of signs and symptoms of relatedness frustration in digital experience. Experiences of relatedness frustration may feel:

- Alienating
- Disconnecting
- Divisive
- Lonely
- Embarrassing/shaming
- Adversarial
- Uncomfortably competitive
- Disrupting of human interaction
- Focused on superficial/extrinsic measures of social connection (status, popularity, image)

When users describe technology use in ways that reference elements such as these, designers can understand this experience in terms of relatedness frustration and look for ways to improve relatedness support.

### 3.5. Managing trade-offs

In considering design for need satisfaction, a question may arise as to trade-offs. Generally, the research conducted on SDT shows that improving support for one psychological need can *increase* satisfaction of other needs. However, there may be times when a frustration at one level may be necessary to prevent more significant or longer lasting frustrations at another level; for example, when honouring someone's autonomy through ensuring informed consent entails an inconvenient extra step. There are no easy answers when it comes to trade-offs. They need to be negotiated by teams, preferably with diverse stakeholder representation and analysis of, not only psychological, but wider ethical and societal implications. The METUX model's "Spheres of Technology Experience" (Peters et al., 2018) provide a framework of analysis for identifying the different and parallel impacts on wellbeing that can occur with technology use, and therefore can assist in analyzing trade-offs. Even where trade-offs are necessary, basic principles of need satisfaction can be applied to optimize implementation (e.g. providing a clear rationale for any inconvenience required to ensure consent).

In addition, all projects face trade-offs with respect to resource allocation. Checking against wellbeing heuristics will inevitably add some overhead to the design process. However, because basic psychological needs are shown to mediate engagement and motivation, as well as wellbeing, any time spent on fulfilling psychological needs will actually contribute to several fundamental HCI goals at once as well as to usability. Moreover, specific scales are available to measure the extent to which a technology satisfies basic psychological needs and these can be used for evaluation, diagnostics and redesign within technology projects (Peters et al., 2018).

### 3.6. Limitations and conclusion

We have drawn parallels to usability heuristics herein, however, the development of usability heuristics unfolded differently. While the guidelines drew on psychology theory, they were primarily derived from hundreds of user tests, and therefore refined against a pool of technology-specific empirical evidence. Today, several decades of proof through practice strengthens their validity. In contrast, at the very birth of a field in wellbeing supportive design, we cannot yet ascribe the same level of assurance to the heuristics proposed herein. Although they are based on evidence collected from decades of research (*some* of which was conducted for technology), there remains some interpretation required for translating existing evidence in other fields to the technology context. Therefore, future testing in practice will almost certainly lead to changes and improvements.

Nevertheless, the fact that these heuristics align with psychological strategies that themselves have been tried and tested in empirical studies across contexts, provides what we believe to be strong reassurance that their implementation stands to significantly improve on the status quo – certainly as compared to applying no wellbeing psychology knowledge at all.

Clearly, optimizing design for a construct as complex and multifaceted as wellbeing will be complex and multifaceted itself, but what the theory and research on Basic Psychological Needs fulfilment from SDT provides is a solid foundation upon which to build our efforts. By informing our design with a better understanding of basic psychological needs and how to support them, we pave the way for more beneficial, more effective, and more fulfilling user experiences.

Thus, we have presented this series of 15 heuristics, together with available evidence and examples of their value, in the hopes of facilitating and accelerating a technology industry move toward design that is more respectful and supportive of psychological wellbeing. We also hope to facilitate more research to substantiate, disprove, or improve the heuristics herein. Our vision is ultimately of a technology industry in which wellbeing support is a fundamental expectation of good design.

## Acknowledgements

The author would like to gratefully acknowledge feedback received from Naseem Ahmadpour and Richard M. Ryan.

## Disclosure statement

No potential conflict of interest was reported by the author(s).

## ORCID

Dorian Peters  http://orcid.org/0000-0002-4767-4198

## References

Adachi, P. J. C., & Willoughby, T. (2011). The effect of video game competition and violence on aggressive behavior: Which characteristic has the greatest influence? *Psychology of Violence*, *1*(4), 259–274. https://doi.org/10.1037/a0024908.

Amershi, S., Weld, D., Vorvoreanu, M., Fourney, A., Nushi, B., Collisson, P., Suh, J., Iqbal, S., Bennett, P. N., Inkpen, K., Teevan, J., Kikin-Gil, R., & Horvitz, E. (2019). Guidelines for human-AI interaction. *Proceedings of the 2019 CHI Conference on Human Factors in Computing Systems*, 1–13. Association for Computing Machinery.

Babich, N. (2020, June 3). *Design patterns: Progressive disclosure for mobile Apps*. Medium.

Brignall, H. (2021). Types of dark patterns. *Dark patterns*. https://www.darkpatterns.org/types-of-dark-pattern.

Brown, K. W., Creswell, J. D., & Ryan, R. M. (2015). *Handbook of mindfulness: Theory, research, and practice*. The Guilford Press.

Brown, K. W., & Ryan, R. M. (2003). The benefits of being present: Mindfulness and its role in psychological well-being. *Journal of Personality and Social Psychology*, *84*(4), 822–848. https://doi.org/10.1037/0022-3514.84.4.822.

Burke, M., Marlow, C., & Lento, T. (2010). Social network activity and social well-being. *Proceedings of the 28th International Conference on Human Factors in Computing Systems*, 1909–1912. Association for Computing Machinery.

Calvo, R. A., Peters, D., Vold, K., & Ryan, R. M. (2020). Supporting human autonomy in AI systems: A framework for ethical enquiry. In C. Burr & L. Floridi (Eds.), *Ethics of digital wellbeing: A multidisciplinary approach*. Springer.

Carpentier, J., & Mageau, G. A. (2013). When change-oriented feedback enhances motivation, well-being and performance: A look at autonomy-supportive feedback in sport. *Psychology of Sport and Exercise*, *14*(3), 423–435. https://doi.org/10.1016/j.psychsport.2013.01.003.

Chen, B., Vansteenkiste, M., Beyers, W., Boone, L., Deci, E. L., Van der Kaap-Deeder, J., Duriez, B., Lens, W., Matos, L., Mouratidis, A., Ryan, R. M., Sheldon, K. M., Soenens, B., Van Petegem, S., & Verstuyf, J. (2015). Basic psychological need satisfaction, need frustration, and need strength across four cultures. *Motivation and Emotion*, *39*(2), 216–236. https://doi.org/10.1007/s11031-014-9450-1.

Church, A. T., Katigbak, M. S., Locke, K. D., Zhang, H., Shen, J., de Jesús Vargas-Flores, J., Ibáñez-Reyes, J., Tanaka-Matsumi, J., Curtis, G. J., Cabrera, H. F., Mastor, K. A., Alvarez, J. M., Ortiz, F. A., Simon, J. Y. R., & Ching, C. M. (2013). Need satisfaction and well-being: Testing self-determination theory in eight cultures. *Journal of Cross-Cultural Psychology*, *44*(4), 507–534. https://doi.org/10.1177/0022022112466590.

Desmet, P., & Pohlmeyer, A. E. (2013). Positive design: An introduction to design for subjective well-being. *International Journal of Design*, *7*(3). https://doi.org/10.1108/10878571011029028

Dwyer, R. J., Kushlev, K., & Dunn, E. W. (2018). Smartphone use undermines enjoyment of face-to-face social interactions. *Journal of Experimental Social Psychology*, *78*, 233–239. https://doi.org/10.1016/j.jesp.2017.10.007.

Fogg, B. J. (2019). *Tiny habits: The small changes that change everything*. Houghton Mifflin Harcourt.

Fong, C., Patall, E., Vasquez, A., & Stautberg, S. (2019). A meta-analysis of negative feedback on intrinsic motivation. *Educational Psychology Review*, *31*(1), 121–162. https://doi.org/10.1007/s10648-018-9446-6

Gaggioli, A., Riva, G., Peters, D., & Calvo, R. A. (2017). Chapter 18—positive technology, computing, and design: Shaping a future in which technology promotes psychological well-being. In M. Jeon (Ed.), *Emotions and Affect in Human Factors and Human-Computer Interaction* (pp. 477–502). Academic Press. https://doi.org/10.1016/B978-0-12-801851-4.00018-5

Goemaere, S., Vansteenkiste, M., & Van Petegem, S. (2016). Gaining deeper insight into the psychological challenges of human spaceflight: The role of motivational dynamics. *Acta Astronautica*, *121*, 130–143. https://doi.org/10.1016/j.actaastro.2015.12.055.

Grinde, B., & Patil, G. G. (2009). Biophilia: Does visual contact with nature impact on health and well-being? *International Journal of Environmental Research and Public Health*, *6*(9), 2332–2343. https://doi.org/10.3390/ijerph6092332.

Hagger, M. S., Koch, S., & Chatzisarantis, N. L. D. (2015). The effect of causality orientations and positive competence-enhancing feedback on intrinsic motivation: A test of additive and interactive effects. *Personality and Individual Differences*, *72*, 107–111. https://doi.org/10.1016/j.paid.2014.08.012

Haidt, J., & Allen, N. (2020). Scrutinizing the effects of digital technology on mental health. *Nature*, *578*(7794), 226–227. https://doi.org/10.1038/d41586-020-00296-x.

Hassenzahl, M., Wiklund-Engblom, A., Bengs, A., Hägglund, S., & Diefenbach, S. (2015). Experience-oriented and product-oriented evaluation: Psychological need fulfillment, positive affect, and product perception. *International Journal of Human-Computer Interaction*, *31*(8), 530–544. https://doi.org/10.1080/10447318.2015.1064664.

IDEO.org. (2015). *Field guide to human centered design* (1st ed.). IDEO.org.

Ingledew, D. K., & Markland, D. (2008). The role of motives in exercise participation. *Psychology & Health*, *23*(7), 807–828. https://doi.org/10.1080/08870440701405704.

Iyengar, S. S., & Lepper, M. R. (2000). When choice is demotivating: Can one desire too much of a good thing? *Journal of Personality and Social Psychology*, *79*(6), 995–1006. https://doi.org/10.1037/0022-3514.79.6.995.

Johnson, L. J., Cronin, L., Huntley, E., & Marchant, D. (2022). The impact of the "RunSmart" running programme on participant motivation, attendance and well-being using self-determination theory as a theoretical framework. *International Journal of Sport and Exercise Psychology*, *20*(1), 102–120. https://doi.org/10.1080/1612197X.2020.1819368.

Kim, E., & Drumwright, M. (2016). Engaging consumers and building relationships in social media: How social relatedness influences intrinsic vs. extrinsic consumer motivation. *Computers in Human Behavior*, *63*, 970–979. https://doi.org/10.1016/j.chb.2016.06.025.

Kori, K., Pedaste, M., Leijen, Ä., & Mäeots, M. (2014). Supporting reflection in technology-enhanced learning. *Educational Research Review*, *11*, 45–55. https://doi.org/10.1016/j.edurev.2013.11.003.

Lonsdale, C., Sabiston, C. M., Raedeke, T. D., Ha, A. S. C., & Sum, R. K. W. (2009). Self-determined motivation and students' physical activity during structured physical education lessons and free choice periods. *Preventive Medicine*, *48*(1), 69–73. https://doi.org/10.1016/j.ypmed.2008.09.013.

Martela, F., Hankonen, N., Ryan, R. M., & Vansteenkiste, M. (2021). Motivating voluntary compliance to behavioural restrictions: Self-determination theory–based checklist of principles for COVID-19 and other emergency communications. *European Review of Social Psychology*, *32*(2), 305–343. https://doi.org/10.1080/10463283.2020.1857082.

Martela, F., & Ryan, R. M. (2020). Distinguishing between basic psychological needs and basic wellness enhancers: The case of beneficence as a candidate psychological need. *Motivation and Emotion*, *44*(1), 116–133. https://doi.org/10.1007/s11031-019-09800-x.

McConnell, A. R., Lloyd, E. P., & Humphrey, B. T. (2019). We are family: Viewing pets as family members improves wellbeing. *Anthrozoös*, *32*(4), 459–470. https://doi.org/10.1080/08927936.2019.1621516.

Naqshbandi, K. Z., Liu, C., Taylor, S., Lim, R., Ahmadpour, N., & Calvo, R. (2020). "I am most grateful." Using gratitude to improve the sense of relatedness and motivation for online volunteerism. *International Journal of Human–Computer Interaction*, *36*(14), 1325–1341. https://doi.org/10.1080/10447318.2020.1746061.

National Disability Authority. (2020). *What is universal design?* Centre for Excellence in Universal Design.

Nielsen, J. (1995). *10 Heuristics for user interface design*. Jakob Nielsen's Alertbox.

Niemiec, C. P., & Ryan, R. M. (2009). Autonomy, competence, and relatedness in the classroom: Applying self-determination theory to educational practice. *Theory and Research in Education*, 7(2), 133–144. https://doi.org/10.1177/1477878509104318.

Peng, W., Lin, J.-H., Pfeiffer, K. A., & Winn, B. (2012). Need satisfaction supportive game features as motivational determinants: An experimental study of a self-determination theory guided exergame. *Media Psychology*, 15(2), 175–196. https://doi.org/10.1080/15213269.2012.673850.

Peters, D. (2014). *Interface design for learning: Design strategies for the learning experience*. New Riders (Voices that Matter).

Peters, D., & Calvo, R. (in press). Self-determination theory and technology design. In R. M. Ryan (Ed.), *Oxford Handbook of Self-determination Theory*. Oxford University Press.

Peters, D., Calvo, R. A., & Ryan, R. M. (2018). Designing for motivation, engagement and wellbeing in digital experience. *Frontiers in Psychology*, 9(MAY), 797. https://doi.org/10.3389/fpsyg.2018.00797.

Quested, E., & Duda, J. L. (2009). Perceptions of the motivational climate, need satisfaction, and indices of well- and ill-being among hip hop dancers. *Journal of Dance Medicine & Science*, 13(1), 10–19.

Rigby, S., & Ryan, R. (2011). *Glued to games: How video games draw us in and hold us spellbound*. Praeger.

Ryan, R. M., & Deci, E. L. (2000). Self-determination theory and the facilitation of intrinsic motivation, social development, and well-being. *The American Psychologist*, 55(1), 68–78. https://doi.org/10.1037/0003-066X.55.1.68.

Ryan, R. M., & Deci, E. L. (2017). *Self-determination theory: Basic psychological needs in motivation, development, and wellness* (1st ed.). The Guilford Press.

Ryan, R. M., & Deci, E. L. (2018). Supporting autonomy, competence, and relatedness: The coaching process from a self-determination theory perspective. In S. English, J. M. Sabatine, & P. Brownell (Eds.), *Professional coaching* (1st ed.). Springer Publishing Company.

Ryan, R. M., & Deci, E. L. (2019). Chapter four - brick by brick: The Origins, development, and future of self-determination theory. In A. J. Elliot (Ed.), *Advances in motivation science* (Vol. 6, pp. 111–156). Elsevier.

Sanders, E, B.-N., & Stappers, P. J. (2014). *Convivial toolbox: Generative research for the front end of design*. BIS Publishers.

Sheldon, K. M., & Filak, V. (2008). Manipulating autonomy, competence, and relatedness support in a game-learning context: New evidence that all three needs matter. *The British Journal of Social Psychology*, 47(Pt 2), 267–283. https://doi.org/10.1348/014466607X238797.

Sheldon, K. M., & Krieger, L. S. (2014). Service job lawyers are happier than money job lawyers, despite their lower income. *The Journal of Positive Psychology*, 9(3), 219–226. https://doi.org/10.1080/17439760.2014.888583.

Sheldon, K. M., Ryan, R. M., Deci, E. L., & Kasser, T. (2004). The independent effects of goal contents and motives on well-being: It's both what you pursue and why you pursue it. *Personality and Social Psychology Bulletin*, 30(4), 475–486. https://doi.org/10.1177/0146167203261883.

Shneiderman, B., Plaisant, C., Cohen, M., & Jacobs, S. (2009). *Designing the user interface: Strategies for effective human-computer interaction* (5th Edition). Addison Wesley.

Stiglic, N., & Viner, R. M. (2019). Effects of screentime on the health and well-being of children and adolescents: A systematic review of reviews. *BMJ Open*, 9(1), e023191. https://doi.org/10.1136/bmjopen-2018-023191.

Szalma, J. L. (2014). On the application of motivation theory to human factors/ergonomics: Motivational design principles for human–technology interaction. *Human Factors*, 56(8), 1453–1471. https://doi.org/10.1177/0018720814553471.

Teixeira, P. J., Marques, M. M., Silva, M. N., Brunet, J., Duda, J. L., Haerens, L., La Guardia, J., Lindwall, M., Lonsdale, C., Markland, D., Michie, S., Moller, A. C., Ntoumanis, N., Patrick, H., Reeve, J., Ryan, R. M., Sebire, S. J., Standage, M., Vansteenkiste, M., … Hagger, M. S. (2020). A classification of motivation and behavior change techniques used in self-determination theory-based interventions in health contexts. *Motivation Science*, 6(4), 438–455. https://doi.org/10.1037/mot0000172.

The Interaction Design Foundation. (2021a). *What are design guidelines?* The Interaction Design Foundation.

The Interaction Design Foundation. (2021b). *What is design thinking?* The Interaction Design Foundation.

van Roy, R., & Zaman, B. (2017). Why Gamification fails in education and how to make it successful: Introducing nine gamification heuristics based on self-determination theory. In M. Ma & A. Oikonomou (Eds.), *Serious games and edutainment applications* (vol. II, pp. 485–509). Springer International Publishing.

Wells, D. L. (2009). The effects of animals on human health and well-being. *Journal of Social Issues*, 65(3), 523–543. https://doi.org/10.1111/j.1540-4560.2009.01612.x.

Wiese, L., Pohlmeyer, A. E., & Hekkert, P. (2020). Design for sustained wellbeing through positive activities—a multi-stage framework. *Multimodal Technologies and Interaction*, 4(4), 71. https://doi.org/10.3390/mti4040071.

Wu, T. (2017). *The attention merchants: The epic scramble to get inside our heads*. Vintage.

Yang, X., & Aurisicchio, M. (2021). Designing conversational agents: A self-determination theory approach. In *Proceedings of the 2021 CHI Conference on Human Factors in Computing Systems* (pp. 1–16). Association for Computing Machinery. https://doi.org/10.1145/3411764.3445445

Yu, S., Levesque-Bristol, C., & Maeda, Y. (2018). General need for autonomy and subjective well-being: A meta-analysis of studies in the US and East Asia. *Journal of Happiness Studies*, 19(6), 1863–1882. https://doi.org/10.1007/s10902-017-9898-2.

Zuboff, S. (2015). Big other: Surveillance capitalism and the prospects of an information civilization. *Journal of Information Technology*, 30(1), 75–89. https://doi.org/10.1057/jit.2015.5.

## About the Author

**Dorian Peters** is Associate Director of the Leverhulme Centre for the Future of Intelligence, University of Cambridge, and Research Associate at Imperial College London. She specialises in design for wellbeing and digital ethics in practice. Her books include Positive Computing (MIT Press) and Interface Design for Learning (Pearson).